\documentclass{osa-article}

\journal{osajournal}

\articletype{Research Article}

\usepackage{lineno}

\begin{document}

\title{Sensitivity of Rydberg-atom receiver to frequency and amplitude modulation of microwaves}

\author{Sebastian Borówka,\authormark{1,2} Uliana Pylypenko,\authormark{1,2} Mateusz Mazelanik,\authormark{1,2} and Michał Parniak\authormark{1,3,*}}

\address{\authormark{1}Centre for Quantum Optical Technologies, Centre of New Technologies, University of Warsaw, Banacha 2c, 02-097 Warsaw, Poland\\
\authormark{2}Faculty of Physics, University of Warsaw, Pasteura 5, 02-093 Warsaw, Poland\\
\authormark{3}Niels Bohr Institute, University of Copenhagen, Blegdamsvej 17, 2100 Copenhagen, Denmark}

\email{\authormark{*}m.parniak@cent.uw.edu.pl} 



\begin{abstract*}
Electromagnetically induced transparency (EIT) in atomic systems involving Rydberg states is known to be a sensitive probe of incident microwave (MW) fields, in particular those resonant with Rydberg-to-Rydberg transitions. Here we propose an intelligible analytical model of Rydberg atomic receiver's response to amplitude- (AM) and frequency-modulated (FM) signals, and compare it with experimental results: we present a setup that allows sending signals with either AM or FM and evaluating their efficiency with demodulation. Additionally, the setup reveals a new detection configuration, using all circular polarizations for optical fields and allowing detection of circularly polarized MW field, propagating colinearly with optical beams. In our measurements we systematically present that several parameters exhibit local optimum characteristics and then estimate these optimal parameters and working ranges, addressing the need to devise a robust Rydberg MW sensor and its operational protocol.
\end{abstract*}

\section{Introduction}

In recent years Rydberg-atom based microwave sensors have attracted attention of many researchers, owing to substantial sensitivity of Rydberg-to-Rydberg microwave (MW) transitions and straightforward measurement scheme. Atomic transitions have been proposed as easily reproducible electric field amplitude standard in the MW regime \cite{Sedlacek12}, and various realizations of MW modulation have led to enhanced detection sensitivity \cite{Liu21,Cai22} (likewise did modulation of optical fields \cite{Li20}), as well as to transmitting both analogue \cite{Deb18,Jiao19,Holloway19AIP,Anderson21,Holloway21} and digital \cite{Meyer18,Holloway19IEEE,Song19,Zou20,Li22,Liu22} information via Rydberg atomic receivers. There were also successful attempts at further characterizing measured MW field properties, such as frequency\cite{Jing20}, phase\cite{Simons19APL,Anderson20,Jing20,Jia21}, polarization\cite{Sedlacek13,Anderson18}, and angle-of-arrival\cite{Robinson21}. Additionally, it was shown that wide, off-resonant spectrum can be covered with Rydberg-based detection\cite{Meyer21,Simons21}. Advances in fabrication of vapor cells, crucial for operation of sensors, have led to miniaturization\cite{Baluktsian10,Daschner12,Simons18}, and the potential for commercialization is considerable. This prospect is very promising, as Rydberg atoms have proved to be a medium suitable for various other applications, such as detection of electric field (of lower frequency than MW)\cite{Miller16,Kumar17n}, where quantum limit has been achieved\cite{Cox18}, and even microwave-to-optical conversion of electromagnetic fields\cite{Han18,Vogt19,Tu22}.

The detection principle has been analyzed for the optimal choice of states employed\cite{Fan15,Chopinaud21}. However, as far as MW modulation is concerned, to our knowledge little has been explored in terms of amplitude- (AM) and frequency-modulation (FM) transfer signal dependence on various parameters (bar the measurements of AM transfer bandwidth\cite{Jiao19} and of optimal coupling field detuning\cite{Zou20} in specific scenarios), although the principle behind detecting both types of modulation has been discussed\cite{Holloway21}. Here we concentrate on this topic, presenting a simple model of modulation transfer and then experimentally verifying parametric dependence of AM and FM transfer, which leads to comparison between both types of modulation, as well as several estimates of optimal working ranges for MW modulation and probe field detuning. The results provided here may prove useful in designing a proper modulation-based Rydberg atomic receiver and its tuning protocol.

Furthermore, almost all of the previous experimental works employ $\pi$ transitions caused by linearly-polarized MW field propagating perpendicularly to laser beams as the working horse of detection setup. This solution requires the MW field to be uniform to take full advantage of the detection sensitivity and is not suitable e.g. for sub-wavelength imaging of material MW response. As a proof-of-concept demonstration we present an alternative design utilizing $\sigma$ transitions from circularly-polarized MW field propagating with laser beams along common axis. This alternative configuration takes advantage from larger dipole moment of MW transition ($2500 a_0 e$ compared to $1900 a_0 e$ of corresponding transition in all-linear $\pi$ configuration). Consequently, we demonstrate detection of MW field propagating through an extremely small surface (${\sim}4000 \mathrm{\mu m}^2$), confirming the prospect of non-disruptive MW field imaging.

\section{Principle and methods}

\subsection{Working principle}

\begin{figure}[h]
\centering\includegraphics[width=\textwidth]{"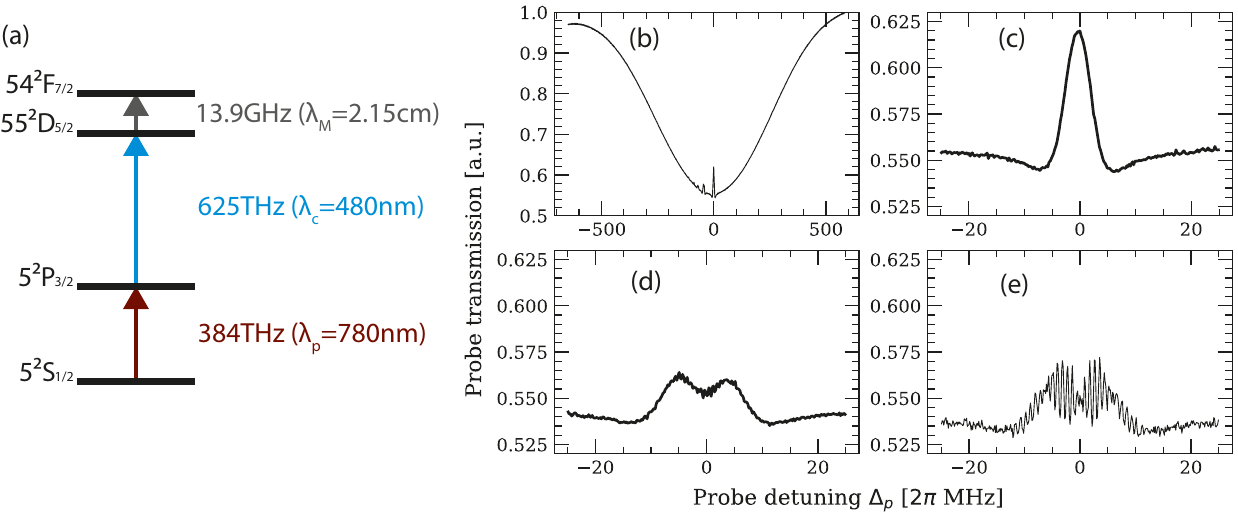"}
\caption{(\textbf{a}) energy level scheme of $^{85}\mathrm{Rb}$, (\textbf{b}) transmission spectrum at $\mathrm{D}_2$ transition with EIT from $5^2\mathrm{P}_{3/2}(\mathrm{F}{=}4) {\rightarrow} 55^2\mathrm{D}_{5/2}$ transition visible in the middle, (\textbf{c}) zoom on the EIT feature, (\textbf{d}) A-T splitting of EIT caused by MW field tuned to $55^2\mathrm{D}_{5/2} {\rightarrow} 54^2\mathrm{F}_{7/2}$ transition, (\textbf{e}) visible wideband FM transferred to probe transmission spectrum. \label{prin}}
\end{figure}

Let us consider 4-level cascade energy level scheme of $^{85}\mathrm{Rb}$, pictured in the Fig. \ref{prin}(a). In the following setup field coupled to $\mathrm{D}_2$ transition between ground state $5^2\mathrm{S}_{1/2}(\mathrm{F}{=}3)$ and excited state $5^2\mathrm{P}_{3/2}$ is employed as a probe $p$. The incident field is absorbed by warm atomic vapors, and the Doppler-broadened spectrum of the transition can be seen in a detuning-transmission plot in the Fig. \ref{prin}(b).

The field coupled to second, $5^2\mathrm{P}_{3/2}(\mathrm{F}{=}4) {\rightarrow} 55^2\mathrm{D}_{5/2}$ transition (coupling field $c$) leads to the emergence of widely-studied electromagnetically induced transparency (EIT) effect\cite{Marangos98,Felischhauer05,Finkelstein22}, which can be seen in the Fig. \ref{prin}(b), as well as zoomed in the Fig. \ref{prin}(c). Strong coupling to the transitions results in rise of dark states, which manifest as transparent resonances in the absorption spectrum. The use of counter-propagating probe and coupling fields can give the advantage of partially canceling Doppler effect in two-photon resonance -- the example in the Fig. \ref{prin}(b,c) shows the resonance narrowed to ${\sim} 5{\cdot} 2 \pi\mathrm{MHz}$ (compared to the full ${\sim} 600{\cdot} 2 \pi\mathrm{MHz}$ absorption profile of the $\mathrm{D}_2$ line).

Then third, $55^2\mathrm{D}_{5/2} {\rightarrow} 54^2\mathrm{F}_{7/2}$ transition is in the interesting MW regime ($f_M = 13.9 \mathrm{GHz}$). As a transition between two Rydberg states, it exhibits very large dipole moment $|d_M| = 2500 a_0 e$ (for comparison the dipole moments for probe $|d_p| = 3 a_0 e$ and coupling field $|d_c| = 0.014 a_0 e$), resulting in great sensitivity. Namely, the MW field amplitude causes Autler-Townes (A-T) splitting of energy levels, that can be read directly as the splitting of EIT feature, as seen in the Fig. \ref{prin}(d). Having taken Doppler effect into consideration, the splitting in the scale of probe detuning $\Delta_p$ can be expressed as $\frac{\lambda_c}{\lambda_p} \Omega_M$, with $\Omega_M$ Rabi frequency being linearly dependent on the MW electric field amplitude $A_M$:
\begin{equation}
    \Omega_M = \frac{d_M}{\hbar}A_M
\end{equation}

When the MW field is amplitude- (AM) or frequency- (FM) modulated, it naturally transfers its modulation to the transmission spectrum of probe field. An exemplary wideband FM spectrum is presented in the Fig. \ref{prin}(e) -- the modulation fringes can be seen due to aliasing effects. These modulations, in terms of their intensity dependence on various parameters, such as probe detuning, are our main focus in this research.

\subsection{Susceptibility model}

Let us consider semi-classical atom-light interaction described with 4-level cascade transition scheme in the Fig. \ref{prin}. Solving GKSL equation (Lindbladian) in rotating wave approximation, in steady state case, with weak probe field assumption, yields us the following formula for probe field's electric susceptibility, in a form of regular nested fraction\cite{Finkelstein22}:
\begin{equation}\label{model}
    \chi = \dfrac{n |d_p|^2}{2 \epsilon_0 \hbar} \dfrac{i}{\gamma_1 - i \delta_1 + \dfrac{(\Omega_c/2)^2}{\gamma_2 - i \delta_2 + \dfrac{(\Omega_M/2)^2}{\gamma_3 - i \delta_3}}}
\end{equation}
where $n$ -- atomic number density, $\delta_{1,2,3}$ -- effective 1,2,3-photon detunings, $\Omega_{c,M}$ -- respective fields' Rabi frequencies, $\gamma_{1,2,3}$ -- respective decay rates of states $5^2P_{3/2}$, $55^2D_{5/2}$ and $54^2F_{7/2}$ (including both natural line widths of these levels and other decay mechanisms, such as transit-time broadening; we assume the decay mechanisms lead to the ground level).

Translation to realistic model, with warm atoms affected by Doppler effect, and parameterized by detuning of lasers, in a realization with probe beam counter-propagating to other fields, requires the following substitutions:
\begin{equation}
    \begin{aligned}
        \delta_1 = \Delta_p + \frac{2 \pi}{\lambda_p} v\\
        \delta_2 = \Delta_p + \Delta_c + \frac{2 \pi}{\lambda_p} v - \frac{2 \pi}{\lambda_c} v\\
        \delta_3 = \Delta_p + \Delta_c + \Delta_M + \frac{2 \pi}{\lambda_p} v - \frac{2 \pi}{\lambda_c} v - \frac{2 \pi}{\lambda_M} v
    \end{aligned}
\end{equation}
where $\Delta_{p,c,M}$ -- proper field detunings, $v$ -- velocity class.

Acquiring full Doppler-broadened spectrum over all velocity classes is done by integration:
\begin{equation}
    \chi_{\mathrm{Dopp}} = \sqrt{\frac{m}{2 \pi k_B T}} \int \chi(v) e^{-\frac{m v^2}{2 k_B T}} dv
\end{equation}
where $m$ -- rubidium atom mass, $T$ -- vapor temperature. 

MW modulation transfer, denoted as $\eta$, for reasonably small modulation frequencies (in particular smaller than $\Omega_M$) can be understood as derivative of susceptibility over Rabi frequency $\Omega_M$ for AM, and derivative over detuning $\Delta_M$ for FM:
\begin{equation}
    \begin{aligned}\label{mod}
        \eta_{\mathrm{AM}} = \frac{\partial}{\partial \Omega_M} \chi_{\mathrm{Dopp}} \\
        \eta_{\mathrm{FM}} = \frac{\partial}{\partial \Delta_M} \chi_{\mathrm{Dopp}}
    \end{aligned}
\end{equation}
The imaginary parts of these modulation transfers (proportional to absorption spectra) as a function of probe detuning are compared to the experimental results in the Fig. \ref{rabi}. These simulated modulation transfer spectra (black) show good agreement with the experimentally obtained (blue, red), in particular concerning the optimal working points. Full model with arbitrarily large or fast modulation calls for a much more complex treatment, for example using the Floquet expansion in the MW frequency, or simulating the full time evolution of the atomic density matrix. This highlights the need for the simpler approach of still considerable predictive power we employ here.

As the parameters introduced are general Rabi frequencies and detunings, the model is applicable not only to the described $^{85}\mathrm{Rb}$ transitions, but also more generally to other configurations involving modulated field detection in 3-step ladder system, as long as the weak probe and counter-propagation conditions are met.


\section{Experimental details}

\begin{figure}[h]
\centering\includegraphics[width=\textwidth]{"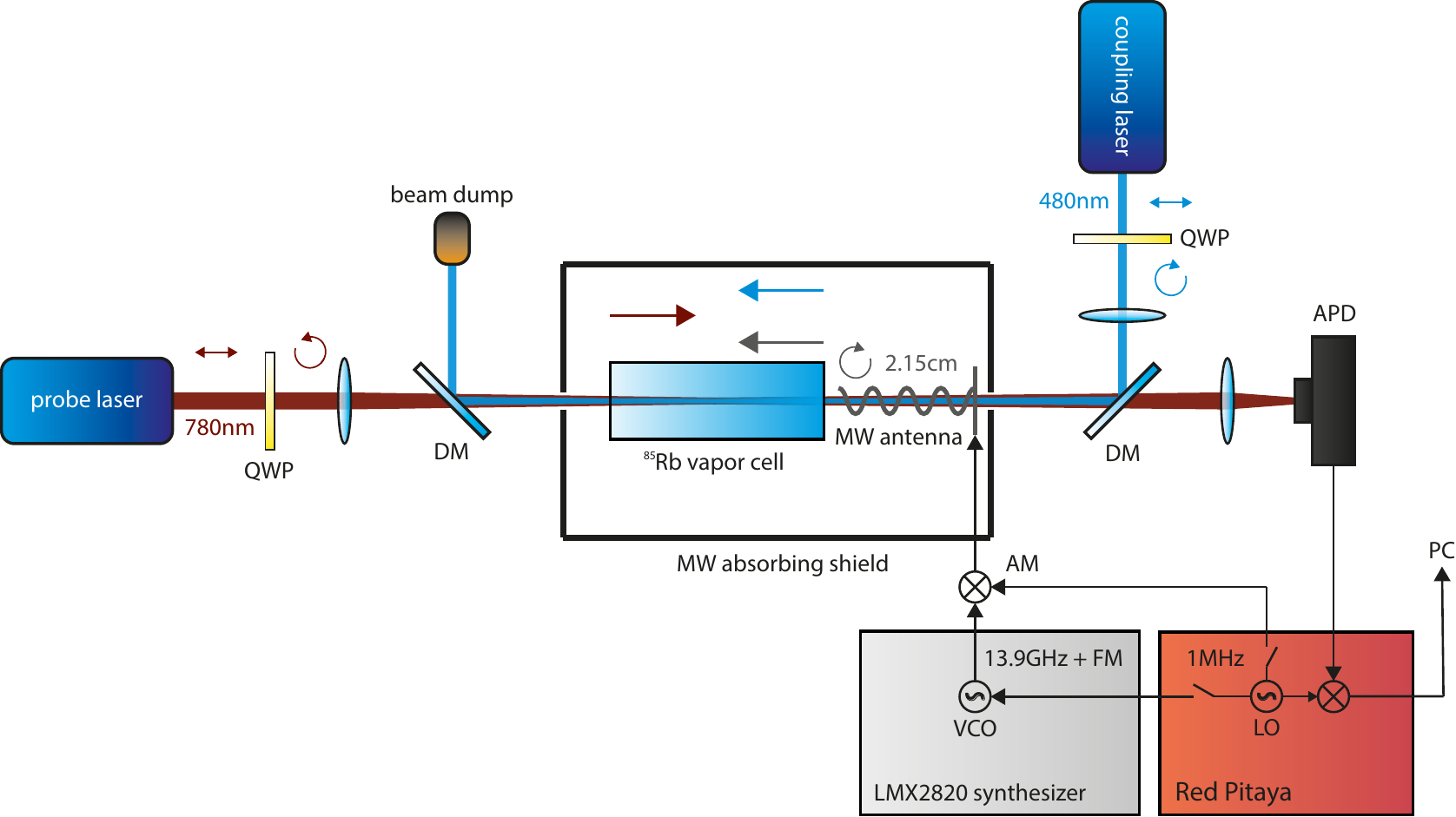"}
\caption{Experimental setup for detecting modulated MW field. QWP -- quarter-wave plate, DM -- dichroic mirror, APD -- avalanche photodiode, VCO -- voltage controlled oscillator, LO -- local oscillator. \label{set}}
\end{figure}

The scheme of the experimental setup used in this research is pictured in the Fig. \ref{set}. The probe (coupling) laser beam of power $P_p = 210 \mathrm{n W}$ ($P_c = 37 \mathrm{mW}$) is focused inside the rubidium vapor cell with waist $w_p = 35 \mathrm{\mu m}$ ($w_c = 40 \mathrm{\mu m}$), resulting in Rabi frequency $\Omega_p = 1.7 {\cdot} 2 \pi \mathrm{MHz}$ ($\Omega_c = 2.9 {\cdot} 2 \pi \mathrm{MHz}$). Both lasers exhibit short-time spectral stability ${<} 100 {\cdot} 2 \pi \mathrm{kHz}$. The counter-propagation of the laser beams results in partial cancellation of Doppler effect and makes it easier to separate the optical fields with dichroic mirrors. The length of the cell is $50 \mathrm{mm}$, however atom-light interaction length can be considered shorter, as Rayleigh lengths for the laser beams are respectively $z_{R,p} = 4.9 \mathrm{mm}$ and $z_{R,c} = 10.5 \mathrm{mm}$.

The MW helical antenna generates RHCP (right-hand circularly polarized) MW field at the $13.9\mathrm{GHz}$ ($2.15\mathrm{cm}$) transition. The antenna is fabricated with a hole in its backplate, so that the beams propagate through it and the whole setup include field propagation along only one axis. For polarization consideration, the coupling beam is made RHCP with a quarter-wave plate and the probe beam (counter-propagating to the other fields) is made LHCP (left-hand circularly polarized). This setup, concerning the energy level structure (Fig. \ref{prin}(a)), assures sign-matched $\sigma$ transitions, which were observed to result in the most prominent EIT and A-T splitting effects in this configuration due to the largest transition dipole moments.

To isolate the setup from spurious MW field, the vapor cell and the antenna are placed inside a shield made from MW-absorbing foam. The apertures where optical beams pass through are considered sub-wavelength to MW field. To avoid interference, all the elements inside the shield, including the cell and its holder, are non-metallic. As the cell is not thermally controlled, the shield also provides thermal insulation assuring relatively constant, $22.5^{\circ}\mathrm{C}$ temperature, which results in constant atomic number density.

MW electrical signal is generated via the LMX2820 phase-locked loop (PLL) frequency synthesizer and can be there regulated in terms of amplitude and frequency. The modulation of MW field is introduced on electrical level, FM is realized with $\Delta f = 1\mathrm{MHz}$ injected into PLL feedback of the synthesizer, and AM is realized with external mixer (Fig. \ref{set}), having the same $1\mathrm{MHz}$ signal in one of the inputs. The strength of modulation (modulation index) is regulated with the amplitude of modulation signal, produced via the Red Pitaya STEMlab 125-14 multifunction measurement tool.

The control over probe detuning is performed with the laser driver scanning of laser current and piezo. After passing through the vapor cell, the probe beam is focused on avalanche photo-diode (APD, Thorlabs APD120A) aperture where transmission is measured. APD signal is then transferred to the STEMlab 125-14, that acts both as analog-to-digital converter and demodulator. The signal is high-pass filtered and quadrature-demodulated, with both operations implemented in the FPGA programmable logic. The data containing raw APD signal, as well as both quadratures of demodulated signal, is then transferred to PC for further processing.

\begin{figure}[h]
\centering\includegraphics[width=12cm]{"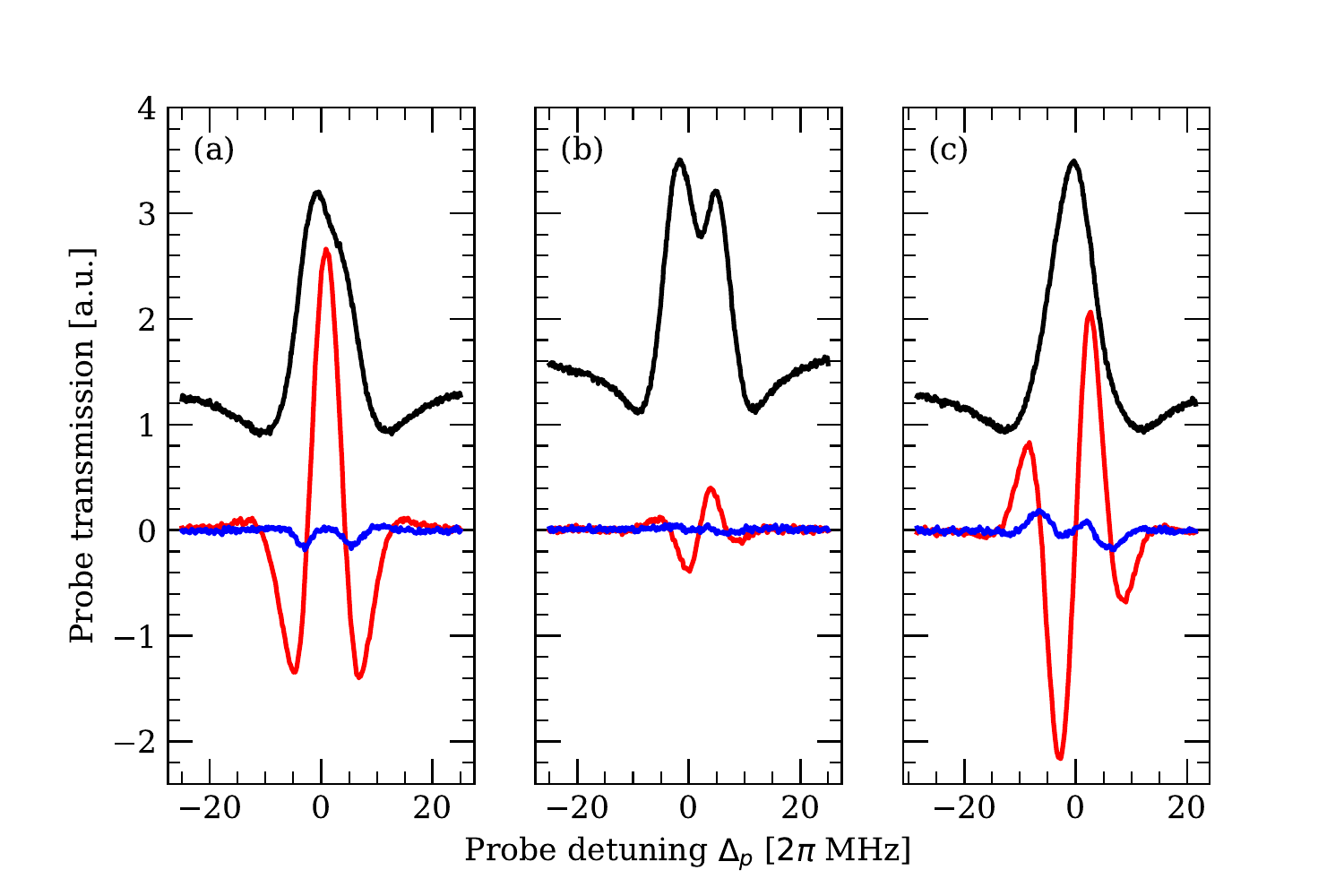"}
\caption{Exemplary measured scan trace with superimposed (\textbf{a}) full amplitude ($h_{\mathrm{AM}} = 1$, $\Omega_M = 7.3 {\cdot} 2 \pi \mathrm{MHz}$), (\textbf{b}) comparable frequency ($h_{\mathrm{FM}} = 1$, $\Omega_M = 8.8 {\cdot} 2 \pi \mathrm{MHz}$),  and (\textbf{c}) wideband frequency modulation ($h_{\mathrm{FM}} = 10$, $\Omega_M = 8.8 {\cdot} 2 \pi \mathrm{MHz}$). Black -- probe transmission spectrum, red -- demodulated signal's proper quadrature, blue -- minimized residual quadrature. \label{exem}}
\end{figure}

Data after demodulation is averaged ($n=10$) for clarity. Demodulation phase is chosen with respect to minimizing one of the quadratures -- the other one is then presented as proper result of modulation transfer parameterized with probe detuning. Exemplary measurements with both quadratures are shown in the Fig. \ref{exem}.

\section{Results and discussion}

\begin{figure}[h]
\centering\includegraphics[width=10cm]{"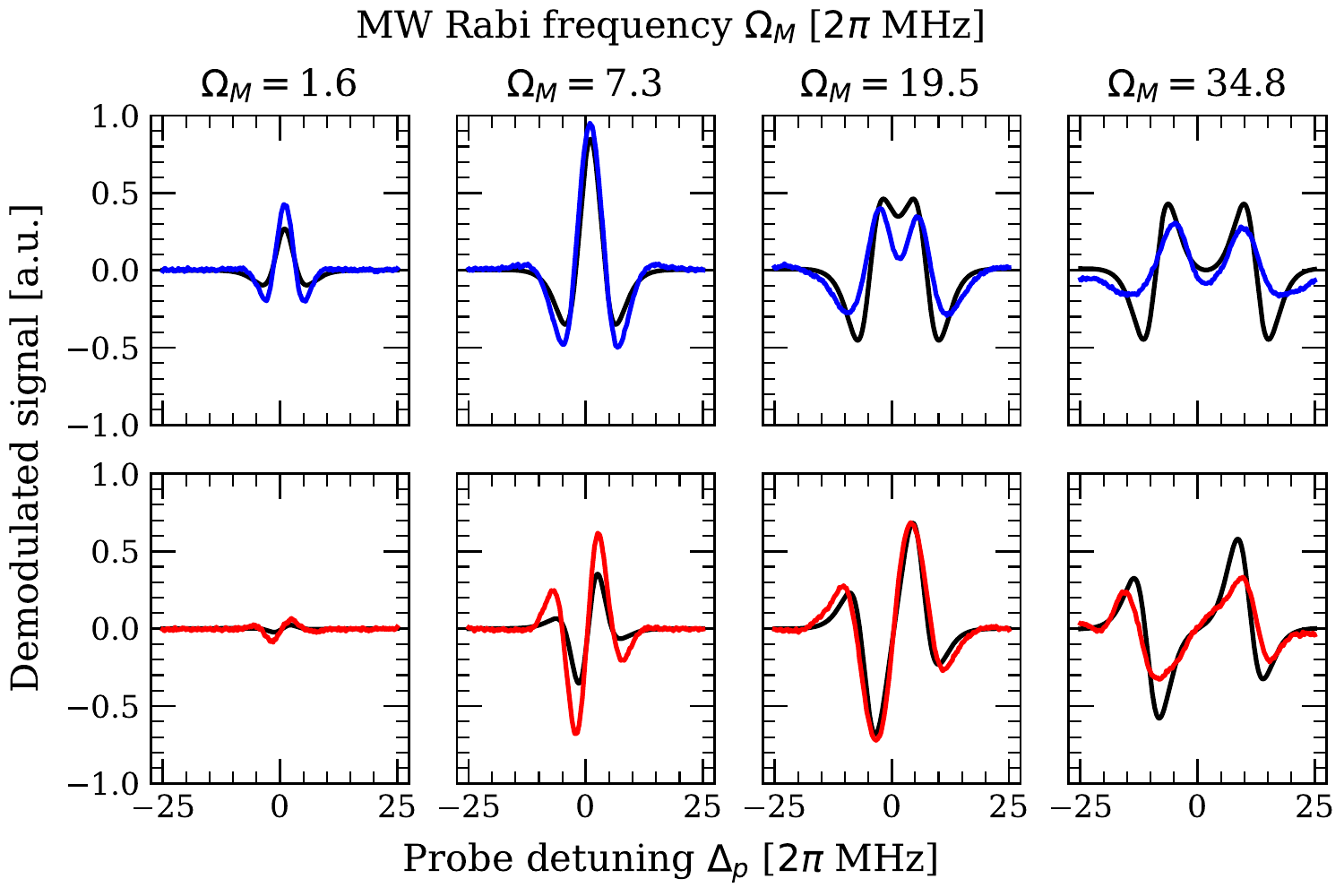"}
\caption{Measured responses to AM (blue, upper row) and wideband ($h_{\mathrm{FM}} = 10$) FM (red, lower row) as a function of probe detuning for various carrier intensities defined with Rabi frequency $\Omega_M$, and corresponding responses simulated in theoretical model (black). \label{rabi}}
\end{figure}

To estimate the MW Rabi frequency, we performed standard measurement of A-T splitting using a non-modulated MW field. Then we translated electrical signal amplitude generated by LMX (checked independently on an electrical spectrum analyzer) to Rabi frequencies. Transit-time broadening caused by atoms flying through the interaction volume was measured by fitting the model to 3-level EIT shape and yielded $\gamma_{\mathrm{t-t}} = 3.9 {\cdot} 2 \pi \mathrm{MHz}$. This seems to be in good agreement with crude yet simple estimation, where $2 \pi \frac{v_{\mathrm{avg}}}{2 w_p} \approx 3.9 {\cdot} 2 \pi \mathrm{MHz}$ for Maxwell's distribution average velocity $v_{\mathrm{avg}}$ in temperature $22.5^{\circ}\mathrm{C}$. As the Rydberg state natural line widths are in kHz range, we assumed $\gamma_2 = \gamma_3 = \gamma_{\mathrm{t-t}}$ in Eq.(\ref{model}) governing our model. As for the decay rate of first excited level, we assumed $\gamma_1 = \frac{\Gamma}{2} + \gamma_{\mathrm{t-t}}$, where $\Gamma = 6.06 {\cdot} 2 \pi \mathrm{MHz}$ is the natural line width of rubidium $D_2$ transition.

If not noted otherwise, all the following results were obtained in case of resonant coupling and MW fields, and the modulation frequency was $\Delta f = 1\mathrm{MHz}$. The modulation frequency $\Delta f$ had to be smaller than MW Rabi frequency $\Omega_M$ for the experiment to operate in the regime where our model applies. On the other hand, $\Delta f$ has to be much larger than laser scanning frequency to avoid undesirable aliasing effects. We decided on $1\mathrm{MHz}$ as it satisfied both conditions and was convenient to work with our electronic setup. However, brief analysis of other frequencies has shown little to no dependence of modulation transfer on modulation frequency in the range $100\mathrm{kHz}$--$10\mathrm{MHz}$.

The results measured in this setup serve as a proof-of-concept demonstration of new $\sigma$ transition based setup, suitable for detection of circularly-polarized MW fields. The weakest measured AM field (Fig. \ref{rabi}, $\Omega_M = 1.6 {\cdot} 2 \pi \mathrm{MHz}$) has the amplitude $A_M = 490\frac{\mathrm{\mu V}}{cm}$, which does not come close to ${<}10\frac{\mathrm{\mu V}}{cm}$ achieved by other groups\cite{Sedlacek12,Kumar17o,Kumar17n,Jing20,Li22}, however, as the area of detection is extremely small, ${\sim}4000 \mathrm{\mu m}^2$ (effective area of probe Gaussian beam cross-section $\pi w_p^2$), in terms of MW photons interacting with atoms we achieve ${\sim}10^9$ microwave photons per second, meaning that very few MW photons need to interact with receiver to result in measurable signal.

\subsection{Comparison between AM and FM}

A quick comparison between amplitude and frequency modulation measurement is shown in the Fig. \ref{exem}. AM signal is more prominent than FM (Fig. \ref{exem}(a,b)) for similar modulation index $h$, which can be understood as a measure of how much energy is transferred to sideband frequencies in the spectral image of modulated signal. In case of AM, modulation index $h_{\mathrm{AM}}=1$ can be considered full modulation, as the electric field decreases to zero. However, for FM modulation index $h_{\mathrm{FM}}>1$ can be achieved and then demodulated signal becomes the most prominent for modulation index $h_{\mathrm{FM}}=10$ (Fig. \ref{exem}(c)). Additionally, the shape of probe transmission spectrum (Fig. \ref{exem}, black) is dependent on modulation type and modulation index.

The interesting aspect of modulated signal detection is optimal probe detuning. It can be seen in the Fig. \ref{exem}, that for AM the optimal detuning is $\Delta_p = 0$, but for wideband FM we measured $\Delta_p = \pm 2.4{\cdot}2 \pi \mathrm{MHz}$. However, this value is dependent on $\Omega_M$ and becomes larger particularly for stronger MW fields, as can be seen in the Fig. \ref{rabi} (lower row, red). The same figure shows that for stronger ($\Omega_M > 8 {\cdot} 2 \pi \mathrm{MHz}$) MW fields AM response changes as well (upper row, blue), with detuning $\Delta_p = 0$ no longer being optimal and the overall response being worse than for FM -- contrary to what was found for smaller MW field intensity. The FM response does not change its shape except for broadening, which may be more useful in potential demodulation tuning protocol. The strongest AM response was estimated for $\Omega_M = 7 {\cdot} 2 \pi \mathrm{MHz}$ and the strongest FM response for $\Omega_M = 14 {\cdot} 2 \pi \mathrm{MHz}$.

In the Fig. \ref{rabi} we also compare results obtained experimentally (blue, red) to simulated theoretical model (black). The model exhibits good prediction of optimal probe detunings and even optimal Rabi frequencies, but slight discrepancies arise when comparing signal transfer amplitude for different Rabi frequencies. These may be attributed to modulation frequency no longer being small enough (compared to $\Omega_M$) for smaller Rabi frequencies, and as for larger Rabi frequencies, additional decoherence from strong A-T splitting may be unaccounted for.

\subsection{FM modulation bandwidth consideration}

\begin{figure}[h]
\centering\includegraphics[width=12cm]{"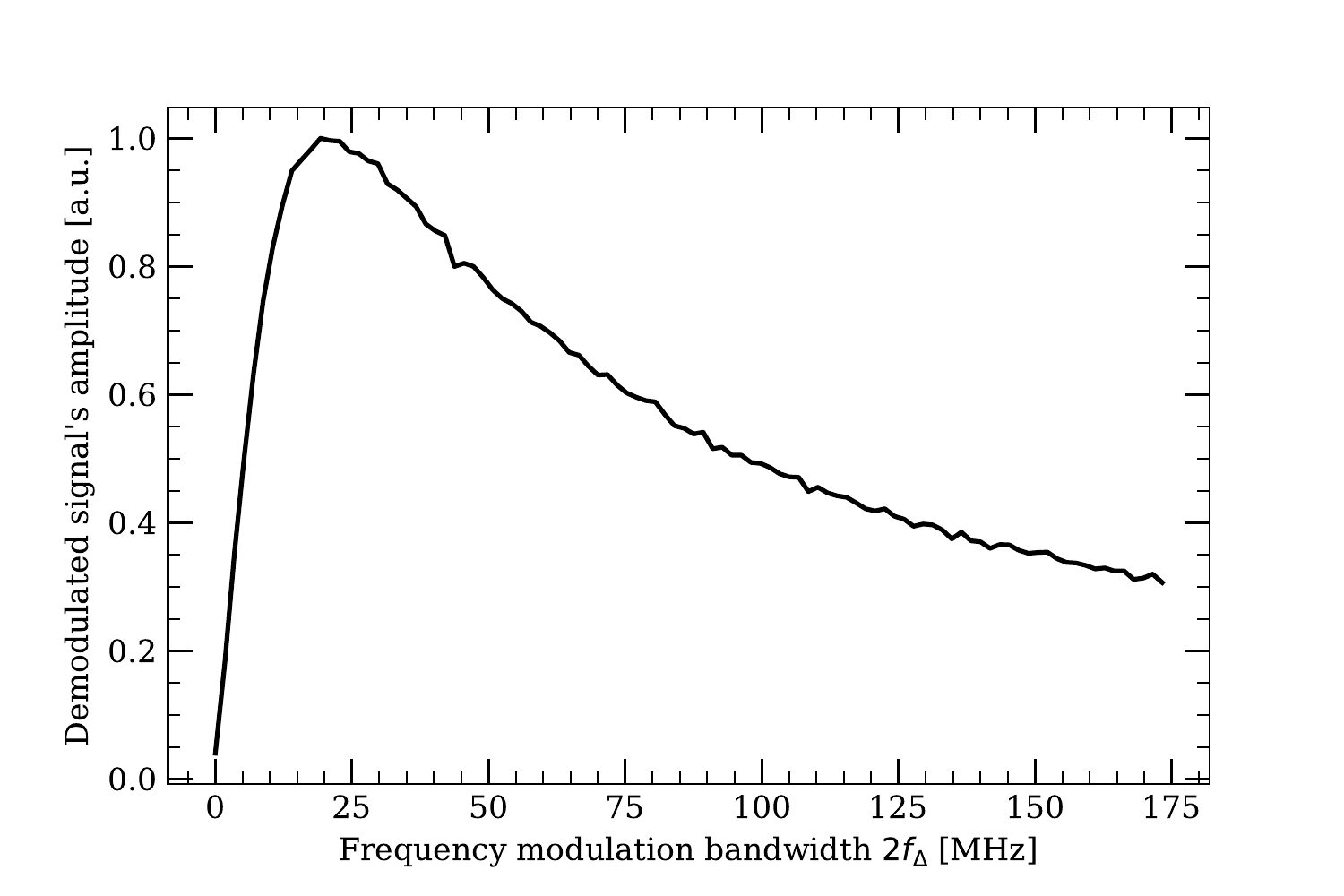"}
\caption{Measured response to FM as a function of modulation bandwidth. The amplitude is normalized to the highest value. Carrier Rabi frequency $\Omega_M = 8.8 {\cdot} 2 \pi \mathrm{MHz}$. \label{band}}
\end{figure}

Having analyzed modulation responses in terms of probe detuning and carrier MW field intensity, let us focus on the bandwidth of FM, the Rydberg atoms acting as receiver can efficiently respond to. The bandwidth of modulation can be understood as twice modulation deviation $2 f_\Delta$, where $f_\Delta = h_{\mathrm{FM}} \Delta f$. We measured FM demodulated signal amplitude for wide range of bandwidths -- the results are shown in the Fig. \ref{band}. We estimated the optimal bandwidth to be $2 f_\Delta = 20 \mathrm{MHz}$, but it can be seen that the modulation signal is received well in a range of bandwidths, having ${>}0.5$ relative efficiency in range $2 f_\Delta = 5$--$100\mathrm{MHz}$. This determines, what bandwidth/modulation index should be chosen to maximize the signal transmission when designing Rydberg receiver and transmission protocol.

\subsection{Wideband FM detuning}

\begin{figure}[h]
\centering\includegraphics[width=12cm]{"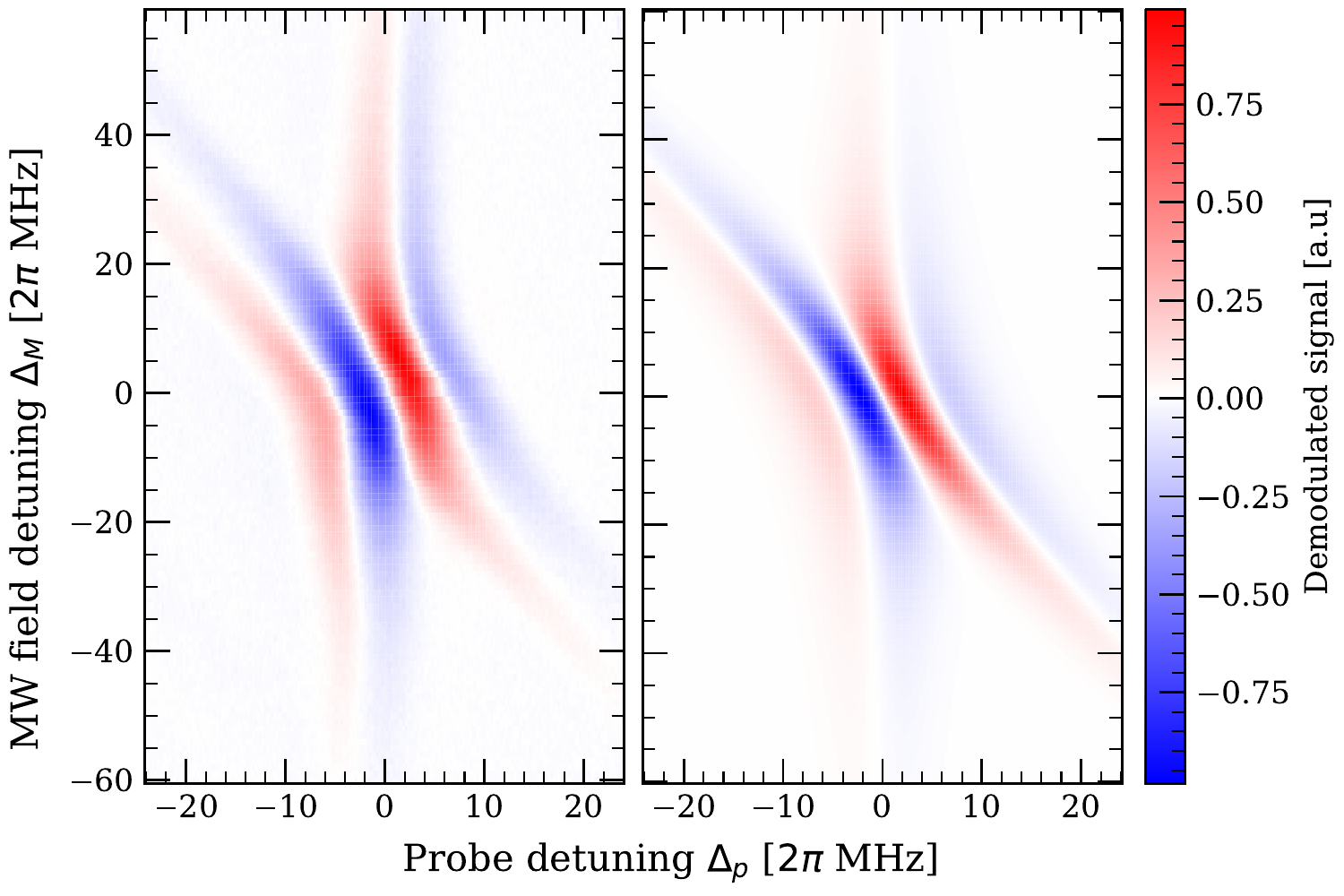"}
\caption{Measured (left) and simulated (right) response to FM as a function of probe detuning and MW field detuning. The amplitude is normalized to the highest value, with respect to zero to show phase transitions. Modulation bandwidth $2 f_\Delta = 20 \mathrm{MHz}$ ($h_{\mathrm{FM}} = 10$), carrier Rabi frequency $\Omega_M = 8.8 {\cdot} 2 \pi \mathrm{MHz}$. \label{map}}
\end{figure}

Now let us consider FM with set bandwidth but carrier frequency detuned from the exact atomic resonance, which addresses the problem of employing various detuned communication channels near the resonant frequency. The demodulated signal amplitude in the function of both probe detuning and MW field detuning are presented as a colormap in the Fig. \ref{map} (left), compared with a colormap obtained from computer simulation (right). The colormaps show two resonances: two-photon resonance and three-photon resonance, consolidating in MW zero-detuning case. It is shown that demodulated signal amplitude weakens for large MW detunings. However, this does not strictly correspond to FM bandwidth, as in this case ${>}0.5$ efficiency is estimated in $\Delta_M = \pm 17 \mathrm{MHz}$ range and our other measurements did not show straightforward dependence on modulation bandwidth.

\section{Summary and perspectives}

The simple model presented in Eqs. (\ref{mod}) has proved to be reliable in terms of predictions of optimal working point dependence on MW field amplitude and detuning, which is particularly important for designing a receiver working in both weak and standard field. 

The estimated optimal working parameters and ranges apply directly only to the employed detection setup. However, as most of them rely directly on the theoretical model, an adequate change of parameters' values in Eq. (\ref{model}), can straightforwardly lead to arriving at optimal working points in other configurations, such as different Rydberg transitions, utilizing different alkali elements and using the well-developed linear polarization of MW field as modulation medium. 

We have demonstrated a setup sensitive to MW fields driving $\sigma$ transitions and shown detection of MW propagating through micrometric surface. Nevertheless, our realization remains relatively simple and can be further improved in demanding applications. For example, highest sensitivity to AM or FM is achieved with additional probe-reference field\cite{Kumar17n,Song19,Liu22} or if atoms are used as a mixer with an external local oscillator (LO)\cite{Kumar17o,Gordon19,Li22}, which increases complexity of the entire setup. Additionally, it was proposed to combine Rydberg sensor with conventional antenna\cite{Simons19IEEE}. Alternative approaches to measurement of MW fields with Rydberg atoms are also possible, e.g. utilizing collective Rabi splitting in a setup with cavity\cite{Yang20}. The measurement scheme can be repeated in cold atoms, trapped in MOT, as MW field amplitude measurements have already been performed\cite{Tanasittikosol11,Liao20}.

\begin{backmatter}
\bmsection{Funding}
Fundacja na rzecz Nauki Polskiej (MAB/2018/4 “Quantum Optical Technologies”); European Regional Development Fund; Narodowe Centrum Nauki (2021/43/D/ST2/03114); Office of Naval Research Global (N62909-19-1-2127).

\bmsection{Acknowledgments}
The “Quantum Optical Technologies” project is carried out within the International Research Agendas programme of the
Foundation for Polish Science co-financed by the European Union under the European Regional Development Fund. This research was funded in whole or in part by National Science Centre, Poland 2021/43/D/ST2/03114. For the purpose of Open Access, the author has applied a CC-BY public copyright license to any Author Accepted Manuscript (AAM) version arising from this submission. MM was also supported by the Foundation for Polish Science via the START scholarship. We thank K. Banaszek and W. Wasilewski for support and discussions.

\bmsection{Disclosures}
The authors declare no conflicts of interest.

\bmsection{Data availability} Data has been deposited at Harvard Dataverse \cite{Data22}.

\end{backmatter}


\bibliography{refs}






\end{document}